\documentclass[floats,aps,showpacs,amssymb,
prd,twocolumn,
,secnumarabic%
,tightenlines%
,nofootinbib]{revtex4}
\usepackage{amssymb}
\usepackage{graphics}
\usepackage{amsmath}
\usepackage{amsfonts}
\usepackage{bm}
\usepackage{graphicx,amssymb,amsmath}

\begin{document}

\title{Probing mixing of photons and axion-like particles by geometric phase}

\author{A. Capolupo, G. Lambiase and G. Vitiello}
\affiliation{Dipartimento di Fisica "E.R. Caianiello" Universit\'a di Salerno,  and INFN - Gruppo Collegato di Salerno, Italy.}
\date{\today}
\def\be{\begin{equation}}
\def\ee{\end{equation}}
\def\al{\alpha}
\def\bea{\begin{eqnarray}}
\def\eea{\end{eqnarray}}

\begin{abstract}

We find that a geometric phase characterizes the phenomenon of mixing of photons with axion-like particles (ALPs). The laboratory observation of such a phase may provide a novel tool able to  detect such a mixing phenomenon.  We show that the geometric phase is  dependent on the axion-like particle mass and coupling constant. We discuss an interferometric experiment able to detect the geometric phase associated to the ALPs-photon mixing.

\end{abstract}


\maketitle

\section{Introduction}

In the physics beyond the Standard Model much attention is devoted to the study of
 ultra-light and weakly coupled particles.
 In particular the axion and axion-like particles (ALPs) and their properties have been throughly analyzed. The study of ALPs starts with the assumption of the existence of the  axion  in order to resolve the strong $CP$ problem in quantum chromodynamics \cite{peccei,peccei1}.
  Such a particle is supposed to be a pseudo-Nambu-Goldstone boson with no electric charge and mass expected to range between $10^{-6}$ and $10^{-2}$ eV, with very small cross-section for strong and weak interactions. An interesting property of the axion and of the ALPs is that
they can couple to an electromagnetic field, so that they should mix  with photons and oscillate in the presence of strong magnetic fields \cite{raffelt}.
Axions and   ALPs are proposed to be natural candidates for cold dark matter. Their existence could be revealed in the anomalous cooling of white dwarfs and the anomalous transparency of the Universe for very high energy gamma rays.

Different kind of terrestrial experiments have attempted to detect axions. The proof of their existence is of course a target  of primary experimental importance.
A small anomalous rotation of the direction of light polarization, propagating through magnetic fields, has been investigated in PVLAS   \cite{PVLAS1}. In CAST,  axions of astrophysical origin are searched by using the Primakoff effect \cite{CAST,CAST1,CAST2} and in ADMX, a strong magnetic field permeating a cold microwave cavity \cite{ADMX,ADMX1} has been utilized to reveal the axion presence. The conversion of photons into axions has been also analyzed in a beam of light passing through an intense magnetic field \cite{11,12,13,16,17}, although without positive results \cite{11,12,13,16,17}.
In  cosmology, a recent analysis of axion dark matter scenario has been carried out in \cite{sikiview1,sikiview2,alessandro}. It has been shown that the axion and ALP contribution to the dark matter can be obtained provided that the axion mass is $m_a = (82.2\pm 1.1)\mu$eV
or $m_a = (76.6\pm 2.6)\mu$eV
(which correspond to the breaking of the Peccei-Quinn symmetry at a scale $f_a =(7.54\pm 0.10)\times 10^{10}$GeV
or $f_a =(8.08\pm 0.27)\times 10^{10}$GeV, respectively).

In recent years, on the other hand, a great attention has been devoted to the study of topologically non-trivial phases, the so-called geometric phases \cite{Berry:1984jv}--\cite{Mostafazadeh1}. These are not reducible to, or derivable from dynamical phases (neither the vice-versa is possible due to their substantial, not slight or negligible difference; the ones cannot be written in terms of the other ones) and have been experimentally
observed in the evolution of many physical systems, from photons in optical fibers \cite{Tomita} and nuclear magnetic resonance \cite{Jones},
to superconducting circuits \cite{Leek,Leek1} and electronic harmonic oscillators \cite{Pechal}. Their possible applications in different field of physics \cite{Blasone:2009xk,Blasone:2009xk1,Capolupo:2013xza,Capolupo:2013xza1,Hu-Yu} have also been studied extensively. Our experience in computing the geometric phase and its consequences in elementary particle physics, in particular in particle mixing and oscillations \cite{Blasone:2009xk,Blas1,Blas2} has inspired us to consider the emergence of topological phases in axion physics. Preliminary results have been presented as an extended abstract in the conference report of Ref.\cite{Capolupo:2015qha}

In this paper we show that the mixing of photon with ALP gives rise indeed to the appearance of a geometric phase.
Specifically, we demonstrate that the geometric phase  represents an observable characterization of the time evolution of photons and ALPs only if  their mixing is present. On the contrary, no geometric phases arise in the absence of mixing and/or vanishing  external magnetic field.
In such cases, the total phase is the dynamical one.
Therefore,  the conclusion is that the geometric phase could be used  as a tool to highlight the generation of axions and ALPs by photons propagating through a strong magnetic field.
One, indeed, could measure the  difference between the geometric phases associated to two rays of light, propagating in two branches of an interferometer, one through a strong magnetic field $B$, the other one passing through a region where $B = 0$.
In particular, we propose to consider a laser beam divided by a beam splitter in two rays of equal intensity, one passing in a region permeated by a strong magnetic field, and the other acting as a reference beam, passing in a region with no magnetic field. The two photon (laser) beams are then recombined together and might interfere.
Indeed, in the presence of photon conversion in ALPs, the beam passing through the region with $B\neq 0$ should have a reduced amplitude and a phase shift with respect to the reference beam \cite{Raffelt-Stodolski}, with the result that there is a reduction in the total intensity of the recombined beam with respect to the original one and an interference pattern due to the difference in the photon phase will appear on the screen.

Here we show that the evolution of photon states, in the presence of magnetic field, contains a phase of geometric origin in addition to the usual dynamical phase derived from Schr�dinger's equation. This  phase results from the geometrical properties of the parameter space of the Hamiltonian of the system. It is gauge invariant and reparametrization invariant. In the particular case of the mixing of photons with ALPs, the geometric phase depends on the mixing angle $\theta$ and therefore on the coupling constant and ALP mass (see below the expression of the mixing angle).

The beam of photons which propagates with no interaction with the magnetic field has geometric phase equal to zero.
The difference between the geometric phases of the beams exiting from the  two branches   is then the manifestation of the occurrence of the ALP-photon mixing phenomenon in the branch where $B\neq 0$.

 The proposal of such an experiment is new, not drawn from previous literature. For example, the experiment presented  in  \cite{tam} is based on a different analysis than the one proposed here. In \cite{tam} it was proposed to use the interferometer solely to reveal the reduction of the power of the laser beam in the output with respect to the input power,  without any consideration concerning the geometrical phases.
 In the seminal paper \cite{Raffelt-Stodolski}  the phase   in the photon-ALP mixing in the presence of
electromagnetic field has been discussed without reference to  the geometric phase, which on the contrary is the main focus of the present paper. Our formalism is fully consistent with the analysis of   \cite{Raffelt-Stodolski}, but it also includes the study of the geometric phase  arising as a consequence of the photon-ALP mixing.
In the Appendix we outline the main contact points of our formalism with \cite{Raffelt-Stodolski}.

We observe that the analysis of the geometric phase applies both to scalar and pseudoscalar ALPs and it can be used   to discriminate between the two cases. We consider the pseudoscalar case.

The paper is structured as follows. In Sec. II we recall the basic formalism describing the axion-photon mixing. In Sec. III we
present the geometric phase characterizing such a phenomenon and study its possible application to the detection of axion and of axion like particles.
Section IV is devoted to conclusions.

\section{axion-photon mixing}

The lagrangian density of the ALP-photon system is given by $L = L_{\gamma} + L_{a} + L_{QED} + L_{a \gamma \gamma} $,
where $L_{\gamma} = - \frac{1}{4} F_{\mu \nu} F^{\mu \nu}$ and $L_{a} = \frac{1}{2} (\partial_{\mu} a \partial^{\mu} a - m_{a}^{2} a^{2})$ are the lagrangian densities of free photon and axion, $L_{QED} =  \frac{\alpha^{2}}{90 m_{e}^{4}}[(F_{\mu \nu } F^{\mu \nu })^{2}+\frac{7}{4}(F_{\mu \nu } \tilde{F}^{\mu \nu })^{2}]$ is the Heisenberg-Euler term due to loop correction in QED  \cite{Heisenberg}-\cite{Dobrich},  which we neglect in the approximation we consider. $L_{a \gamma \gamma} $ is given by
\bea\label{1}
L_{a \gamma \gamma} = \dfrac{g_{a \gamma \gamma}}{4} \, a \, F^{\mu \nu}\tilde{F}^{\mu \nu}\,,
\eea
and it represents the interaction of two photons with the axion pseudoscalar field $a$ in the presence of a magnetic field \cite{PDG}. Eq.(\ref{1}) can be written as $L_{a \gamma \gamma} = -g_{a \gamma \gamma} a\, {\bf E}  \cdot {\bf B} $. Here  $\tilde{F}_{\mu \nu } = \frac{1}{2}\epsilon_{\mu \nu \rho \sigma}F^{\rho \sigma} $ is the  dual electromagnetic tensor,
$g_{a\gamma \gamma}\equiv g \equiv g_{\gamma} \alpha/ \pi f_{a} $ is the axion-photon coupling with dimension of inverse energy,  with $g_{\gamma}$  of the order of the unity, $\alpha =1/137$,  $f_{a}$ the decay constant of ALPs.
By means of such an interaction, axion and photon can mix each other in a background magnetic field.

We consider a monochromatic laser beam splitted in two beams, one propagating through a region permeated by a magnetic field and the other traversing a region with $B =0$. The two beams exiting from the two branches are then recombined together. We assume that the beam propagating in the presence of the magnetic field has the direction  along the z-axes. Besides  the $L_{QED}$ term, we also neglect the birefringence of fluids in a transverse magnetic field (Cotton-Mouton effect). With this simplifications, choosing the y-axis along the projection of $\textbf{B}$  perpendicular to the z-axes, the photon polarization state $\gamma_{x} = \gamma_{\bot}$ decouples and the propagation equations can be written as
\bea
\left(\omega - i \partial_{z} + \textit{M}  \right)\left(
                                                \begin{array}{c}
                                                  \gamma_{\|} \\
                                                  a \\
                                                \end{array}
                                              \right) = 0\,,
\eea
with $\textbf{B} = \textbf{B}_{T}$, purely transverse field and mixing matrix $\textit{M}$ given by
\bea
\textit{M} = -\frac{1}{2\omega}\, \left(
               \begin{array}{cc}
                 \omega_{P}^{2} & - g \omega B_{T} \\
                 - g \omega B_{T} & m_{a}^{2} \\
               \end{array}
             \right).
\eea
Here $\omega_{P}$ is the plasma frequency and $m_{a}$ is the axion mass. The matrix \textit{M} can be diagonalized by a rotation to primed fields
\bea
\left(
  \begin{array}{c}
    \gamma^{\prime}_{\|}(z) \\
    a^{\prime}(z) \\
  \end{array}
\right) = \left(
            \begin{array}{cc}
              \cos \theta & \sin \theta \\
              -\sin \theta & \cos \theta \\
            \end{array}
          \right)
          \left(
            \begin{array}{c}
              \gamma_{\|}(z) \\
              a(z) \\
            \end{array}
          \right)\,.
\eea
The mixing angle is $\theta = \frac{1}{2} \arctan \displaystyle{\left(\frac{2 g \omega B_{T}}{m_{a}^{2}-\omega^2_P} \right)}$,
moreover $\gamma^{\prime}_{\|}(z) = \gamma^{\prime}_{\|}(0) e^{- i \omega_{\gamma} z}$, $a^{\prime} (z) = a^{\prime} (0) e^{- i \omega_{a} z}$
where
\[
\omega_{\gamma} = \omega + \Delta_{-}\,, \quad
\omega_{a} = \omega+ \Delta_{+}\,,
\]
and
\[
\Delta_{\pm} = - \frac{\omega_{P}^{2}+ m_{a}^{2}}{4 \omega} \pm \frac{1}{4 \omega} \sqrt{(\omega_{P}^{2}- m_{a}^{2})^{2} + (2 g \omega B_{T})^2}\,.
\]
In this way, the photons  can be converted into ALPs in the branch of the interferometer in which $B \neq 0$. The conversion of photons in ALPs  manifests itself in a reduced amplitude of that beam  and in the phase shift with respect to the reference beam.

For the mixing components one has
\bea
\left(
  \begin{array}{c}
    \gamma_{\|}(z) \\
    a(z) \\
  \end{array}
\right) = \emph{M}(z)
          \left(
            \begin{array}{c}
              \gamma_{\|}(z) \\
              a(z) \\
            \end{array}
          \right)\,,
\eea
where
\begin{widetext}
\bea\label{M1}
\emph{M}(z) =  \left(
            \begin{array}{cc}
              \cos \theta & -\sin \theta \\
              \sin \theta & \cos \theta \\
            \end{array}
          \right)
          \left(
            \begin{array}{cc}
              e^{-i \omega_{\gamma} z}   & 0 \\
            0 &  e^{-i \omega_{a} z}   \\
            \end{array}
          \right)
          \left(
            \begin{array}{cc}
              \cos \theta &  \sin \theta \\
              -\sin \theta & \cos \theta \\
            \end{array}
          \right)
          \,.
\eea
Using Eq.(\ref{M1}),   $\gamma_{\|}(z)$ is obtained as 
\bea\label{gamma}
\gamma_{\|}(z) = \left[e^{-i \omega_{\gamma} z} \cos^{2} \theta + e^{-i \omega_{a} z} \sin^{2} \theta \right] \gamma_{\|}(0) + \cos\theta \sin\theta  \left[e^{-i \omega_{\gamma} z} - e^{-i \omega_{a} z}\right] a(0)
\eea
and similar for $a(z)$. By assuming that the contribution coming from the amplitude $a(0)$ of the axion field be negligible  with respect to the $\gamma$-term in the r.h.s. of Eq. (\ref{gamma}),  we have
\bea\label{gamma1}
\gamma_{\|}(z) =\emph{M}_{11}(z)\gamma_{\|}(0)  = \left[e^{-i \omega_{\gamma} z} \cos^{2} \theta + e^{-i \omega_{a} z} \sin^{2} \theta \right] \gamma_{\|}(0)\,.
\eea
 We observe that the correspondence between  the notation of Ref.~\cite{Raffelt-Stodolski} and our notation is  $\Delta_{\|} = - \frac{\omega_{P}}{2 \omega}$, $\Delta_{a}= -\frac{m_{a}}{2 \omega}$, $\Delta^{\prime}_{\|} = \Delta_{+}$, $\Delta^{\prime}_{a} = \Delta_{-}$.  By following \cite{Raffelt-Stodolski},  we may
neglect the  phase $e^{-i (\omega + \Delta_{\|})z }$ which is common to $\gamma_{\|}(z)$, $\gamma_{\bot}(z)$ and $a_{\|}(z)$; the matrix $\emph{M}(z)$ becomes
\bea\label{M}
\emph{M}(z) =  \left(
            \begin{array}{cc}
              \cos \theta & -\sin \theta \\
              \sin \theta & \cos \theta \\
            \end{array}
          \right)
          \left(
            \begin{array}{cc}
              e^{-i(\Delta^{\prime}_{\|}-\Delta_{\|})z}   & 0 \\
            0 &  e^{-i(\Delta^{\prime}_{a}-\Delta_{\|})z}   \\
            \end{array}
          \right)
          \left(
            \begin{array}{cc}
              \cos \theta &  \sin \theta \\
              -\sin \theta & \cos \theta \\
            \end{array}
          \right)\,
\eea
%
and the states (\ref{gamma1}) is then
\bea\label{gamma3}
\gamma_{\|}(z) = \left[e^{-i (\Delta_{\|}^{\prime}-\Delta_{\|}) z} \cos^{2} \theta + e^{-i (\Delta_{a}^{\prime}-\Delta_{\|}) z} \sin^{2} \theta \right] \gamma_{\|}(0)\,.
\eea
\end{widetext}

In the following Section we focus our attention on the geometric phase generated  in the laser beam which has been  propagating through the branch with the magnetic field. The photon beam carrying such a phase, exiting from such a branch, will then interfere with the photon reference beam.
 
 In closing this Section, we observe that it has been shown \cite{van} that the transition rate can be enhanced by filling the conversion region with a buffer gas.
This induces an effective photon mass $m_{\gamma}$.
The plasma frequency
$\omega_{P}=(4 \pi \alpha N_{e} /m_{e} )^{1/2}$, where $N_{e}$ denotes the electron density, plays the role of $m_{\gamma}$ \cite{van}.

\section{Geometric phase and axion-photon mixing}

In order to show that a geometric phase arises in the evolution of the ALP-photon system, i.e. in the laser beam passing through the magnetic field, we use the definition of Mukunda-Simon (geometric) phase \cite{Mukunda}. Let us therefore briefly introduce such a definition.

Such a phase, derived within a kinematical approach, is associated to any open curve of unit vectors in Hilbert space.
For a quantum system whose state vector $|\psi(s)\rangle$ belongs to a curve $\textit{C}$, with $s$ real parameter such that $s \in [s_1, s_2]$, the Mukunda--Simon phase is defined as \cite{Mukunda}:
\bea\label{geofase}
\Phi (\textit{C}) = \arg \langle \psi(s_{1})| \psi(s_{2} )\rangle - \Im\int_{s_{1}}^{s_{2}}\langle \psi(s)|\dot{\psi}(s)\rangle d s\,,
\eea
where the dot denotes the derivative with respect to the real parameter $s$.
The right side of Eq.(\ref{geofase}) is the first term in Eq.(2.11) of Ref.~\cite{Mukunda}.
 In Eq.(\ref{geofase}), $\arg \langle \psi(s_{1})| \psi(s_{2} )\rangle$ represents the total phase, while $ \Im\int_{s_{1}}^{s_{2}}\langle \psi(s)|\dot{\psi}(s)\rangle d s$ is the dynamical one.
The difference between the two, if not zero, is by definition the geometric phase.

To our knowledge the geometric phase has not been considered in \cite{Raffelt-Stodolski} and in other existing literature relative to axion and ALPs.
The definition in  Eq.(\ref{geofase}) can be applied to the photons interacting with the magnetic field. In this case, the geometric phase is given by
\bea
\Phi_{ \gamma}(z)  =  \arg \langle  \gamma_{\|} (0 )| \gamma_{\|}(z)\rangle\, -\, \Im \int_{0}^{z}  \langle   \gamma_{\|}(z^{\prime})| \dot{\gamma}_{\|}(z^{\prime})\rangle  d z^{\prime}\,,
\eea
with $| \gamma_{\|}(z )\rangle$ the state of the photons. For $\gamma_{\|}(z )$, given by Eq.(\ref{gamma3}), we obtain
\bea\label{MSphase}\nonumber
\Phi_{ \gamma}(z) & = & \arg \left[e^{-i (\Delta_{\|}^{\prime}-\Delta_{\|}) z} \cos^{2} \theta + e^{-i (\Delta_{a}^{\prime}-\Delta_{\|}) z} \sin^{2} \theta \right]
\\\nonumber
& + & \left[ (\Delta_{\|}^{\prime}-\Delta_{\|}) \cos^{4}\theta + (\Delta_{a}^{\prime}-\Delta_{\|}) \sin^{4}\theta \right] z
\\
& - & \frac{1}{4}\sin^{2}2\theta  \cos \left(\frac{\Delta_{osc} z}{\cos 2 \theta} \right),
\eea
with $\Delta_{osc} \equiv \Delta_{\|} - \Delta_{a}$  (as in the notation of \cite{Raffelt-Stodolski}).
 Let us introduce the notation $A = 1 + \theta^{2} (\cos \Delta_{osc}z -1 )$ and $B= -\theta^{2} ( \Delta_{osc}z - \sin \Delta_{osc}z )$. We then consider Eq.(\ref{MSphase}) at the second order in the mixing angle $\theta$ and obtain: 

\begin{widetext}

\bea \nonumber i)~~~ \Phi_{ \gamma}(z) &=& \frac{\pi}{2} + \theta^{2} (\Delta_{osc}z + \cos \Delta_{osc}z)\,,  \qquad \qquad ~~~~~~~~~{\rm for} ~ A=0, ~B>0\,, \\
\nonumber ii)~~~ \Phi_{ \gamma}(z) &=& \frac{3\pi}{2} + \theta^{2} (\Delta_{osc}z + \cos \Delta_{osc}z)\,,  \qquad \qquad ~ ~~~~~~~{\rm for} ~ A=0, ~B<0 \,,\\
\nonumber iii)~~~ \Phi_{ \gamma}(z) &=& \theta^{2} (2\Delta_{osc}z -\sin \Delta_{osc}z + \cos \Delta_{osc}z)\,, \qquad ~~~~~{\rm for} ~ A>0, ~B\geq 0\,,\\
\nonumber iv)~~~  \Phi_{ \gamma}(z) &=& \theta^{2} (2\Delta_{osc}z -\sin \Delta_{osc}z + \cos \Delta_{osc}z) +2 \pi\,, \quad ~{\rm for} ~  A>0, ~B< 0\,,\\
\nonumber v)~~~ \Phi_{ \gamma}(z) &=& \theta^{2} (2\Delta_{osc}z -\sin \Delta_{osc}z + \cos \Delta_{osc}z) +  \pi\,.  ~~~~~~{\rm for} ~   A<0,  ~\forall B\,. \eea

\end{widetext}

The non-vanishing value of the geometric phase $\Phi_{ \gamma}$ indicates univocally the existence of the ALP-photon mixing phenomenon.
Indeed, $\Phi_{ \gamma}(z) = 0 $ when the the magnetic field is zero and the mixing angle $\theta $ is equal to zero. This is the case of the reference beam. See the Appendix for a comparison of Eq.(\ref{MSphase}) with the phase definition in
 \cite{Raffelt-Stodolski}.

  The geometric phase  is thus calculated for the mixed photon state.
It can be detected in the end crossing point of the laser beams (in which the magnetic field is zero),
by analyzing the interference between the laser beam passing through the magnetic field and  the  reference beam.
Indeed, the geometric phase accumulated by the photons during their paths through the magnetic field  survives when the photons exit the magnetic field  and interfere with the reference beam. 

Notice that, the geometric phase can be detected in interferometric experiments when the dynamical phase is much smaller than the geometric one.
In general, paths of slightly different lengths can be chosen so that the geometric phase dominates over the dynamical one.
In the particular case of the mixing, the phase
acquired in the branch with $B \neq 0$ will have both a geometrical and a dynamical component. The total phase is
\bea\nonumber
\Phi_{ TOT}^{B \neq 0}(z) &=& \Phi_{ \gamma}(z)
\\\nonumber
& - &
 \left[ (\Delta_{\|}^{\prime}-\Delta_{\|}) \cos^{4}\theta + (\Delta_{a}^{\prime}-\Delta_{\|}) \sin^{4}\theta \right] z
\\\nonumber
& + & \frac{1}{4}\sin^{2}2\theta  \cos \left(\frac{\Delta_{osc} z}{\cos 2 \theta} \right)
\\
& \simeq &  \Phi_{ \gamma}(z) + \theta^{2} (\Delta_{osc}z + \cos \Delta_{osc}z)
\eea
where the second term on the r.h.s. is the dynamical phase in the presence of mixing.
On the contrary, the phase of the beam passing in the $B = 0$ branch contains only the dynamical phase
\bea\nonumber
\Phi_{ TOT}^{B = 0}(z) &=&   (\Delta_{\|}^{\prime}-\Delta_{\|})  z
= - \frac{\Delta_{osc}z}{2} \left(1 - \frac{1}{\cos 2\theta} \right)
\\
& \simeq & \theta^{2}  \Delta_{osc}z \,.
\eea

 By accurately designing  the interferometer, the length of the two branches  can be  linked by the relation
\bea\label{time}
l_{B=0} = \frac{ \Delta_{osc} l_{B \neq 0} + \cos (\Delta_{osc} l_{B \neq 0} )}{\Delta_{osc}}
\eea

Building the interferometer in this way, once the two laser  beams are recombined at the detector, the phase difference will be given by the geometric component.
 The contributions of the dynamical phases of the two beams, in this case, will indeed compensate each other since $\Delta_{osc} l_{B=0} =   \Delta_{osc} l_{B \neq 0} + \cos (\Delta_{osc} l_{B \neq 0} ) $ and the phase difference $\Delta \Phi$ will in practice coincide with the
geometric phase $\Phi_{ \gamma}$ induced by the mixing, i.e.  $\Delta \Phi = \Phi_{ \gamma}$.

The interferometer can be built in a way that the beams traverse the branches back and forth many times
before recombining at the detector. This can be achieved by  suitable use of the  mirrors.
In this way, the   phase will add many times.
In principle, one can obtain effective branches of arbitrary length.
In the particular case in which the photons interact with the magnetic field in a region of length the order of the axion-photon oscillation length
$ l_{osc} = \frac{2 \pi}{\Delta_{\|} - \Delta_{a}}$,
the geometric phase (\ref{MSphase}) reduces to the Berry-like phase \cite{Berry:1984jv},
 \begin{equation}\label{berryphase}
\Phi_{ \gamma}(z = l_{osc})  = 2 \pi \sin^{2}\theta \,.
\end{equation}
Eq. (\ref{berryphase}) can be then rewritten for the n-cycle
case as
\bea
 \Phi_{ \gamma}(n l_{osc})  = 2 n \pi \sin^{2}\theta \approx  2 n \frac{g^{2}B^{2}\omega^{2}}{\left(m_{a}^{2}-\omega^{2}_{P} \right)^{2}}\,.
 \eea
A Fabry-Perot cavity can be also used in order to enhance the power of the laser light beam.

For our estimations, we shall utilize Eq.(\ref{MSphase}) and use the following values of the parameters entering the above equations:
magnetic field $B = 10 \, \rm{T}\simeq 1.95\times 10^{-15} \,\rm{GeV}^2$, energy $\omega = 10$ eV, coupling constant $g \in [10^{-16} , 10^{-10}]$GeV$^{-1}$, axion mass $m_{a} \in [10^{-6} , 10^{-2}]$ eV and plasma frequency $\omega_{P} = 0.9 m_{a}$.

 If the experiment runs for the time interval $t = 24$ hours,
we can obtain observable values of the phases $\Delta \Phi $  as function of the masses and of $g$.
Notice that, as observed in \cite{Raffelt-Stodolski} only the photon component of the beam is reflected by mirrors, since any physical mirror, for what concern our discussion, is transparent to axions. After $N$ reflections the compound effect is $\phi(L) = N \phi(  l)$, with $L = N l$ and $l$   the distance between the reflecting mirrors \cite{Raffelt-Stodolski}.

By   fixing the  values of the ALP mass and  varying the values of $g$, we obtain the following results for the geometric phase,

\begin{widetext}
 \begin{center}

\bea 
\nonumber i)~~~  m_{a} & \sim & 10^{-2}   eV, \qquad   g \in [10^{-11} , 10^{-10}]  GeV^{-1}  \quad \rightarrow \quad \Phi_{ \gamma}(N l) \in [10^{-9} , 10^{-7}]\pi \,, 
\\
\nonumber ii)~~~   m_{a} & \sim & 10^{-3}   eV, \qquad  g \in [10^{-12} , 10^{-10}]  GeV^{-1} \quad \rightarrow \quad  \Phi_{ \gamma}(N l) \in [10^{-8} , 10^{-4}]\pi \,,
\\
\nonumber iii)~~~  m_{a} & \sim &  10^{-4}   eV, \qquad  g \in [10^{-14} , 10^{-10}]  GeV^{-1} \quad  \rightarrow \quad  \Phi_{ \gamma}(N l) \in [10^{-8} , 1.6]\pi \,,
\\
\nonumber iv)~~~  m_{a} & \sim &  10^{-5}  eV, \qquad  g \in [10^{-15} , 10^{-10}]  GeV^{-1} \quad  \rightarrow \quad \Phi_{ \gamma}(N l) \in [10^{-6} , 0.9]\pi \,,
\\
\nonumber v)~~~  m_{a} & \sim &  10^{-6}   eV, \qquad  g \in [10^{-16} , 10^{-10}]  GeV^{-1} \quad \rightarrow \quad  \Phi_{ \gamma}(N l) \in [10^{-4} , 0.6]\pi \,.
 \eea

\end{center}

\end{widetext}

These values of $\Phi_{ \gamma}(N l)$ show that the geometric phase is  dependent on the value of the axion mass and on the coupling constant $g$.
Such values are   detectable, therefore the geometric phase could represent a new tool  to reveal the existence of ALPs and also to provide a region of exclusion for the parameter characterizing the ALP physics.

One could wander that the running time  $t = 24 \, h $ is too long in order for our system to survive to decoherence phenomenon. However, as in the interferometry  experiment proposed by Tam and Yang \cite{tam}, one may consider running time as long as several days ($10$ days, $240 h$, in \cite{tam}) provided that one uses laser beams, as it is indeed in our proposed experiment. On the contrary, this would not be the case by considering non-coherent light, which would be trapped in between mirrors in optical cavity systems for a time interval  $\ll 1\, sec$.

The present technologies allow for phase measurements within a precision of $ \sim 10^{-9}$ \cite{stefano,stefano2,stefano3}.
Then, the values of the geometric phase here obtained are detectable,  provided a sufficent intensity of the signal and
signal-to-noise ratio is obtained (see below).

Summing up, the presence of a non-trivial geometric phase of the photon demonstrates the presence of axion-photon mixing; it is  highly sensible to the axion mass and $g$
and it allows to derive
a region of exclusion for the  values of the parameters characterizing the axion physics.

We also observe that the so-called Schottky (or shot) noise, associated to the particle nature of light,  affects the sensibility of the interferometer. In the standard case, the shot noise for  photons from laser light represents the relative fluctuations in the photon number per unit time and it can be significant when the number of photons is small \cite{Schottky}.
Its magnitude increases as the square root of the expected number $N$ of photons. The signal-to-noise ratio is given by
$r = N/\sqrt{N}=\sqrt{N}\,$ \cite{Schottky}.
In the case of the axion-photon mixing, the probability of their conversion is $(g B L)^{2}/4\,,$ which is obtained for axion mass  much smaller than the energy of photons.  The signal-to-noise ratio is then given by $ (g B L)^{2} N / 4 \sqrt{N} \,$.
However, in a squeezed state the number of photons measured per unit time can have fluctuations smaller than the square root of the expected number of photons counted in that period of time \cite{caves}. This implies the possibility to increase the sensibility of the interferometer if squeezed light is used in the experiment
\cite{caves1,caves2}.

A nonclassical noise reduction of almost $13$ dB below vacuum noise has been observed \cite{Eberle}.
Thus, the squeezed light technique has a great application potential.
Injected into an interferometer, the quantum noise reduction corresponding to an
increase of a factor 10 in laser light power is possible.
The use of Fabry-Perot cavity (FPC) in the arms of the interferometer can further increase the sensitivity of the measurements.
  We stress that a tipical precision in phase measurement, of the order of $10^{-9}$, does not in general guaranties the detectability of the phase, which also depends on the intensity of the signal and in turn on the  (small values of the) mixing angle, i.e. on $g$. In this connection, in designing the interferometer,  we expect that also in our case the strategy outlined in \cite{tam} might be adopted in order to boost the sensitivity; namely one can incorporate in our setup optical delay lines or    Fabry-Perot cavities in order to enhance the signal by a factor given by the number of times the laser beam is reflected back and forth in the FPC. By using the estimation made in \cite{tam}, one might improve the signal-to-noise ratio even by ten    dB provided one uses  squeezed laser beams as already mentioned above.

\section{Conclusions}

We have shown that the axion-photon mixing exhibits a characterizing geometric phase.
The existence of the geometric phase for photons propagating in a magnetic field
signals unequivocally  the occurrence of the axion-photon mixing and may provide a novel tool in the detection of such a  phenomenon.
The relation between axion mass and coupling constant could be
tested  in a dedicated interferometer experiment.

For appropriate values of the photon energies $\omega$ and magnetic fields $B$, the geometric phase could provide a parameter window where searching for ALP contribution to cold dark matter and possible contributions of particle mixing phenomena to the Universe dark energy \cite{Capolupo:2006et,Capolupo:2006et1,Capolupo:2006et2,Capolupo:2006et5}.


\section*{Conflict of Interests}
The authors declare that there is no conflict of interests
regarding the publication of this paper.

\section*{Acknowledgements}
Partial financial support from MIUR and INFN is acknowledged.
G.L. thanks the ASI (Agenzia Spaziale Italiana) for partial support through the contract ASI number I/034/12/0.

\section*{Appendix}

We compare the geometric phase for mixed photons derived in Eq.(\ref{MSphase}) with the phase presented in \cite{Raffelt-Stodolski}.
Considering the matrix (\ref{M}) and assuming $a(0)\approx 0$,  the field $\gamma_{\|}(z)$ is given by Eq.(\ref{gamma1}),
here reported again for convenience (see also Eq.(\ref{gamma3}))
\bea\label{gamma2}\nonumber
\gamma_{\|}(z)=\emph{M}_{11}(z)\gamma_{\|}(0) = \left[e^{-i \omega_{\gamma} z} \cos^{2} \theta + e^{-i \omega_{a} z} \sin^{2} \theta \right] \gamma_{\|}(0).
\eea

By following Ref.\cite{Maiani}, one can define the phase $\phi(z)$ for photons in the presence of axions as
\bea\label{fase} \nonumber
\phi(z) = - \Im \emph{M}_{11}(z) = \left[e^{-i \omega_{\gamma} z} \cos^{2} \theta + e^{-i \omega_{a} z} \sin^{2} \theta \right],
\eea
which for small mixing angle is given by
\bea \nonumber
\phi(z) =\theta^{2} (\Delta_{osc}z - \sin \Delta_{osc}z).
\eea
Such a phase is the one reported in \cite{Raffelt-Stodolski} (Eq.(22) of \cite{Raffelt-Stodolski}).
It coincides, up the second order (for $A>0$ and $B\geq 0$, cf. Section 3), with the total phase
$\phi_{TOT}(z) = \arg \langle  \gamma_{\|} (0 )| \gamma_{\|}(z)\rangle\, = \arg \emph{M}_{11}(z)$,
which is also computed in the present paper.
As remarked in the text (cf. Section 3) the geometric  phase is by definition the difference between the total phase
and the dynamical one and represents a topological invariant; it is invariant under gauge  and reparametrization transformations.
The phase presented in \cite{Raffelt-Stodolski} is thus a different quantity with respect to the
geometric phase discussed in the present paper and shown to be nonvanishing.

\end{document}